\documentclass[aps,prd,amsmath,showpacs,nofootinbib]{revtex4}
\usepackage{graphicx}

\newcommand{\beq}{\begin{equation}}
\newcommand{\eeq}{\end{equation}}

\newcommand{\bea}{\begin{eqnarray}}
\newcommand{\eea}{\end{eqnarray}}

\newcommand{\bs}{\begin{subequations}}
\newcommand{\es}{\end{subequations}}

\newcommand{\Ref}[1]{(\ref{#1})}

\begin{document}

\title{Slowly rotating wormholes: the first order
approximation}

\author{P.E. Kashargin$^a$}

\author{S.V. Sushkov$^{a,b}$}
\email{sergey_sushkov@mail.ru; sergey.sushkov@ksu.ru}
\affiliation{$^a$Department of General Relativity and Gravitation,
Kazan State University, Kremlevskaya str. 18, Kazan 420008,
Russia} %
\affiliation{$^b$Department of Mathematics, Tatar State
University of Humanities and Education, Tatarstan str. 2, Kazan
420021, Russia}

\date{\today}

\begin{abstract}
We discuss a solution describing a rotating wormhole in the theory
of gravity with a scalar field with negative kinetic energy. To
solve the problem we use the assumption about slow rotation. The
role of a small dimensionless parameter plays the ratio of the
linear velocity of rotation of the wormhole's throat and the
velocity of light. The rotating wormhole solution is constructed
in the framework of the first order approximation with respect to
the small parameter. We analyze the obtained solution and study
the motion of test particles and the propagation of light in the
spacetime of rotating wormhole.
\end{abstract}

\maketitle

\section{Introduction}
Wormholes are usually defined as topological handles in spacetime
linking widely separated regions of a single universe, or
``bridges'' joining two different spacetimes
\cite{MorTho,VisserBook}. As is well-known \cite{HocVis}, they can
exist {only if} their throats contain an exotic matter which
possesses a negative pressure and violates the null energy
condition. The search of realistic physical models providing the
wormhole existence represents an important direction in wormhole
physics. Various models of such kind include scalar fields
\cite{Ell,Bro,scalarfields}; wormhole solutions in semi-classical
gravity \cite{semiclas}; solutions in Brans-Dicke theory
\cite{Nan-etal}; wormholes on the brane \cite{wormholeonbrane};
wormholes supported by matter with an exotic equation of state,
namely, phantom energy \cite{phantom}, the generalized Chaplygin
gas \cite{chaplygin}, tachyon matter \cite{tachyon}, etc
\cite{footnote,review}.

It is worth to notice that the most investigations deal with
spherically symmetric wormholes because of their high symmetry. At
the same time it would be important and interesting with the
physical point of view to study rotating wormholes. Some general
properties of rotating wormholes have been discussed in the
literature \cite{Teo,Kha,BerHib,Kuh,Kim}. However, so far there
are no exact solutions describing such objects. The aim of this
paper consists in constructing such the solutions.

As is well-known (see \cite{Ell,Bro}) a scalar ghost, i.e. a
scalar field with negative kinetic energy can support static
spherically symmetric wormholes. Moreover, such the wormholes are
stable against linear spherically symmetric perturbations
\cite{stability}. In this paper we look for rotating wormholes
supported by the scalar field with negative kinetic energy. To
solve the problem we suppose that a wormhole is very slowly
rotating and construct a solution in the framework of the first
order approximation with respect to a small parameter
characterizing the velocity of rotation.

The paper is organized as follows. In the section \ref{genform} we
give some general formulas and write down the field equations. A
static spherically symmetric wormhole is briefly discussed in the
section \ref{statwh}. In the section \ref{rotwh} we formulate a
condition of slow rotation and introduce a small parameter
characterizing the velocity of rotation. In the framework of the
first order approximation with respect to the small parameter we
construct a solution describing the rotating wormhole. The
solution is analyzed in the section \ref{analysis}. In the section
\ref{testpart} we study the motion of test particles and the
propagation of light in the spacetime of rotating wormhole. The
summary of results obtained is given in the section \ref{conc}.

\section{General formulas\label{genform}}
Consider the theory of gravity with a scalar field $\phi$
describing by the action
\beq\label{action}
S=\int d^4x\sqrt{-g}\left[R + (\nabla\phi)^2\right],
\eeq
where $g_{\mu\nu}$ is a metric, $g=\det(g_{\mu\nu})$, $R$ is the
scalar curvature, and
$(\nabla\phi)^2=g^{\mu\nu}\phi_{,\mu}\phi_{,\nu}$ is the kinetic
term. Throughout the paper we use units $G=c=1$ and the signature
$(-+++)$. For this signature the sign `$+$' before the kinetic
term corresponds to negative kinetic energy, hence $\phi$ is a
{\em ghost}.

Varying the action \Ref{action} with respect to $g_{\mu\nu}$ and
$\phi$ yields Einstein equations and the equation of motion of the
scalar field, respectively: \bs\label{sys}
\bea\label{einstein}
&&R_{\mu\nu}=-\phi_{,\mu}\phi_{,\nu},\\
\label{eqmo} &&\nabla^{\alpha}\nabla_{\alpha}\phi=0. \eea \es

In the paper we will search for solutions of the system \Ref{sys}
describing rotating wormholes. A spacetime with the stationary
rotation possesses the axial symmetry. As is known (see, e.g.
\cite{Hartle}) a general axially symmetric metric can be given in
the following form:
\beq\label{metric}
ds^2=-Adt^2+Bdr^2+K^2[d\theta^2+\sin^2\theta(d\varphi-\omega
dt)^2],
\eeq
where $A$, $B$, $K$, $\omega$ are functions of $r$, $\theta$. The
function $\omega$ has an explicit physical sense; it represents an
angular velocity of rotation in a point $(r,\theta)$. The
requirement of finiteness of the angular momentum $J$ measured by
a distant observer yields the following asymptotical condition for
$\omega$ \cite{Land}:
\beq\label{as}
\omega=\frac{2J}{r^3}+O(r^{-4})\quad {\rm at}\quad r\to\infty.
\eeq
Also, requiring that a spacetime should be asymptotically flat we
have $A\to1$, $B\to1$, and $K^2\to r^2$ at $r\to\infty$.

Equations of the system \Ref{sys} written down for the metric
\Ref{metric} are partial differential equations of second order
for five functions $A$, $B$, $K$, $\omega$, and $\phi$. Solving
these equations in a general form is rather complicated
mathematical problem. Further, to simplify the problem we restrict
ourselves by the case of slow rotation.

\section{Static spherically symmetric wormhole\label{statwh}}
In order to formulate a condition of the slow rotation we first of
all discuss a static spherically symmetric case. The static
spherically symmetric solution in the theory of gravity with the
ghost scalar field was first found by Ellis \cite{Ell} and,
independently, Bronnikov \cite{Bro}. This solution can be
represented as follows (see \cite{SusKim})
\beq\label{stat.m}
ds^2=-e^{2u}dt^2+e^{-2u}[dr^{2}+(r^{2}+r_0^2)(d\theta^2+\sin^2\theta
d\varphi^2 )],
\eeq
\beq\label{stat.phi}
\phi(r)={\frac{(m^2+r_0^2)^{1/2}}{2\pi^{1/2}\, m}}\,u(r),
\eeq
where the radial coordinate $r$ varies from $-\infty$ to $\infty$,
$m$ and $r_0$ are free parameters, and
\beq\label{u}
u(r)=\frac{m}{r_0}\left(\arctan\frac{r}{r_0}-\frac{\pi}{2}\right).
\eeq
Taking into account the following asymptotical behavior: $
e^{2u}|_{r\to\infty}=1-\frac{2m}{r}+O(r^{-2})$, and $
e^{2u}|_{r\to-\infty}=e^{-2\pi
m/r_0}\left(1-\frac{2m}{r}\right)+O(r^{-2})$, we may see that the
spacetime with the metric \Ref{stat.m} possesses by two
asymptotically flat regions. These regions are connected by the
throat whose radius corresponds to the minimum of the radius of
two-dimensional sphere, $R^2(r)=e^{-2u(r)}(r^2+r_0^2)$. The
minimum of $R(r)$ is achieved at $r_{th}=m$. The value
$R_{th}=R(r_{th})$ is called the radius of wormhole throat. For
$m=0$ the metric \Ref{stat.m} takes the especially simple form:
\beq\label{MTwh}
ds^2=-dt^2+dr^2+(r^2+r_0^2)(d\theta^2+\sin^2\theta d\varphi^2 ),
\eeq
It is worth noting that the metric \Ref{MTwh} was proposed {\em a
priori} by Morris and Thorne in the pioneering work \cite{MorTho}
as a simple example of the wormhole spacetime metric.

\section{Rotating wormhole\label{rotwh}}
Now consider rotating wormholes. For this aim we take the wormhole
metric in the form \Ref{metric}, where $r\in(-\infty,+\infty)$.
Assume that the throat of wormhole corresponds to the value
$r=r_{th}$. The throat's radius we define as
$r_0=K|_{r=r_{th},\theta=\pi/2}$. Also we define the value
\begin{equation}\label{Omega}
\Omega=\omega|_{r=r_{th},\theta=\pi/2},
\end{equation}
being the equatorial angular velocity of rotation of the wormhole
throat. Without loss of generality we may suppose that $\Omega>0$,
i.e. the throat is rotating in the positive direction. Assume now
that the following condition is fulfilled:
\begin{equation}\label{medl}
\Omega r_0\ll c,
\end{equation}
where $c$ is the velocity of light.  The condition \Ref{medl}
means that the linear velocity of rotation of the throat is much
less than $c$. Further we will consider an {\em approximation of
slow rotation} with the small dimensionless parameter
$\alpha\equiv\Omega r_0/c$. In this approximation components of
the metric \Ref{metric}, describing the rotating wormhole, should
just slightly differ from respective components of the static
metric \Ref{stat.m}. Following the procedure given in
\cite{Hartle}, we represent the metric functions $A$, $B$, $K$,
$\omega$ and the field $\phi$ as an expansion in terms of powers
of $\alpha$. At the same time it is necessary to emphasize that
the expansion for the angular velocity should only contain odd
powers of $\alpha$, while the expansions of other functions should
only contain even powers:
\begin{eqnarray}\label{razl-w}
\omega&=&\alpha \omega^{(1)}+O(\alpha^{3}),\\
A&=&A^{(0)}+\alpha^{2}A^{(2)}+O(\alpha^{4}),\\
B&=&B^{(0)}+\alpha^{2}B^{(2)}+O(\alpha^{4}),\\
K&=&K^{(0)}+\alpha^{2}K^{(2)}+O(\alpha^{4}),\\
\phi&=&\phi^{(0)}+\alpha^{2}\phi^{(2)}+O(\alpha^{4}).
\label{razl-phi}
\end{eqnarray}
The zero order functions $A^{(0)}$, $B^{(0)}$, $K^{(0)}$,
$\phi^{(0)}$ correspond to the unperturbed static solution
(\ref{stat.m}), \Ref{stat.phi} as follows
\begin{equation}\label{abkp}
A^{(0)}=e^{2u},\quad B^{(0)}=e^{-2u},\quad
K^{(0)}=e^{-u}(r^2+r_0^2)^{1/2},\quad
\phi^{(0)}={\frac{(m^2+r_0^2)^{1/2}}{2\pi^{1/2}\, m}}\,u(r),
\end{equation}

Further we restrict our consideration to the first order
approximation. In this approximation the functions $A$, $B$, $K$,
$\phi$ remain to be unperturbed and are determined by the
expressions \Ref{abkp}, while the angular velocity $\omega$, being
initially equal to zero, takes the form
$\omega=\alpha\omega^{(1)}$. The metric of rotating wormhole now
reads
\beq\label{firstorder.m}
ds^2=-e^{2u}dt^2+e^{-2u}dr^{2}+e^{-2u}(r^{2}+r_0^2)[d\theta^2+\sin^2\theta
(d\varphi^2-2\alpha\omega^{(1)} dtd\varphi)].
\end{equation}

Substituting the expressions (\ref{razl-w}-\ref{razl-phi}) into
the field equations (\ref{sys}) and keeping terms up to the first
order we obtain the only one nontrivial equation for
$\omega^{(1)}$:
\begin{equation}\label{equ}
-\frac{1}{\sin^3\theta} \partial_{\theta}
[\sin^3\theta\,\partial_{\theta}\omega^{(1)}]=
(r^2+r_0^2)\partial_{r}^2\omega^{(1)}+4(r-m)\partial_{r}\omega^{(1)}.
\end{equation}
To solve this equation we use the method of partition. Namely,
substituting the representation $\omega^{(1)}=\Theta(\theta)W(r)$
into Eq. \Ref{equ} and separating variables yields two ordinary
differential equations for the functions $\Theta(\theta)$ and
$W(r)$:
\begin{equation}\label{Theta}
\frac{1}{\sin^3\theta}
[\sin^3\theta\,\Theta_\theta]_\theta=\lambda\Theta,
\end{equation}
\begin{equation}\label{W}
(r^2+r_0^2)W''+4(r-m)W'=\lambda W,
\end{equation}
where the prime means the derivative with respect to $r$.
Analyzing Eq. \Ref{W} it is easy to obtain an asymptotic for $W$
at $|r|\to\infty$:
\begin{equation}\label{asW}
W= C_1 r^{[-3+\sqrt{9+4\lambda}\,]/2}+C_2
r^{[-3-\sqrt{9+4\lambda}\,]/2},
\end{equation}
where $C_1$, $C_2$ are constants of integration. Comparing two
asymptotics, \Ref{asW} and \Ref{as}, we find that $\lambda=0$. In
this case the solution of Eq. \Ref{Theta} being regular in the
interval $[0,\pi]$ is $\Theta={\rm const}$ (without loss of
generality one may set $\Theta=1$). A general solution of Eq.
\Ref{W} for $\lambda=0$ reads
\begin{equation}
W(r)=C_1+C_2e^{4u(r)}\left[1+\frac{4m(r+2m)}{r^2+r_0^2}\right],
\end{equation}
where the function $u(r)$ is given by \Ref{u}. In order to fix the
values of constants of integration $C_1$, $C_2$ we should use
boundary conditions. First, let us take into account that  $W\to0$
at $r\to\infty$ (see \Ref{as}). From this we find
\begin{equation}\label{c}
C_1=-C_2.
\end{equation}
Second, consider the condition \Ref{Omega} determining the
equatorial angular velocity of rotation of the wormhole throat.
Now it takes the form $W(0)=1/r_0$ (remember that $c=1$). The last
relation together with the condition \Ref{c} let us fix the values
of constants $C_1$, $C_2$. Finally, the expression for the angular
velocity $\omega$ in first order with respect to $\alpha$, i.e.
$\omega=\alpha\omega^{(1)}$, is the following:
\begin{equation}\label{sol}
\omega(r)=\alpha\omega^{(1)}=\frac{\alpha}{r_0[1-e^{-2\pi
m/r_0}(1+8m^2r_0^{-2})]}
\left\{1-e^{4u(r)}\left(1+\frac{4m(r+2m)}{r^2+r_0^2}\right)\right\}.
\end{equation}

\section{Analysis of solution\label{analysis}}  %
\begin{figure}[h]
 \centerline{\includegraphics[width=6cm]{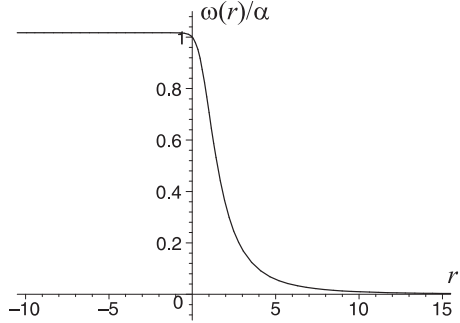}} \caption{The graph of
 $\omega(r)/\alpha$ for $m=1$, $r_0=1$. \label{fig1}}
\end{figure}

Examine the solution \Ref{sol} for $\omega(r)$ in more details. In
the figure \ref{fig1} the graph of the function $\omega(r)$ is
shown.
In the limit $r\to+\infty$ one has $\omega\to0$. With the physical
point of view this means that the coordinate frame of a distant
observer does not rotate. To define the angular momentum $J$ of
the rotating wormhole we consider the expansion of $\omega(r)$
given by Eq. \Ref{sol} at large $r$:
\begin{equation}
\omega(r)=\frac{8\alpha m(r_0^2+4m^2)}{3r_0[1-e^{-2\pi m/r_0}
(1+8m^2r_0^{-2})]}\,{r^{-3}}+O(r^{-4}) .
\end{equation}
Comparing this expression with the asymptotic \Ref{as} we find
\begin{equation}\label{J}
J=\frac{4\alpha m(r_0^2+4m^2)}{3r_0[1-e^{-2\pi m/r_0}
(1+8m^2r_0^{-2})]}.
\end{equation}
Note that the denominator $1-e^{-2\pi m/r_0}(1+8m^2r_0^{-2})$ in
Eq. \Ref{J} is strictly positive if $m>0$ (the case $m=0$ will be
considered separately), hence the angular momentum $J$ is also
positive.

Now consider the second asymptotical region $r=-\infty$ connecting
with the region $r=+\infty$ by the wormhole throat. The function
$\omega(r)$ in the limit $r\to-\infty$ has the following
asymptotical form
\begin{equation}
\omega(r)=\frac{\alpha(1-e^{-4\pi m /r_0})}{r_0[1-e^{-2\pi m/r_0}
(1+8m^2r_0^{-2})]}-\frac{8\alpha m(r_0^2+4m^2) e^{-4\pi m
/r_0}}{3r_0[1-e^{-2\pi m/r_0}
(1+8m^2r_0^{-2})]}\,{|r|^{-3}}+O(|r|^{-4}) .
\end{equation}
It is seen that the value of the angular velocity $\omega(r)$ at
$r\to-\infty$ tends to the constant $\omega_0$, where
\begin{equation}
\omega_0\equiv\omega(-\infty)=\frac{\alpha(1-e^{-4\pi m
/r_0})}{r_0[1-e^{-2\pi m/r_0} (1+8m^2r_0^{-2})]}.
\end{equation}
This means that the coordinate frame, being non-rotating in the
region $r=+\infty$, receives a constant rotation with the angular
velocity $\omega_0$ in the region $r=-\infty$.

In the case $m=0$ the expression \Ref{sol} for $\omega(r)$ is
simplified and takes the following form:
\begin{equation}\label{omegam0}
\omega(r)=\frac{\alpha}{r_0}\left[1-\frac2\pi\frac{rr_0}{r^2+r_0^2}
-\frac2\pi\arctan\frac{r}{r_0}\right].
\end{equation}
The respective angular momentum $J$ now reads
\begin{equation}
J=\frac{2\alpha r_0^2}{3\pi}.
\end{equation}

\section{Motion of test particles\label{testpart}}
In this section we will study a motion of test particles in the
rotating wormhole spacetime with the metric \Ref{firstorder.m}.
Thereinafter we will consider only the equatorial plane motion and
set $\theta=\pi/2$. To derive the equations of motion we use the
Hamilton-Jacobi method \cite{Land}. The Hamilton-Jacobi equation
for a particle with the mass $\mu$ reads
\begin{equation}
g^{\alpha\beta}\frac{\partial S}{\partial
x^{\alpha}}\frac{\partial S}{\partial x^{\beta}}+\mu^2=0.
\end{equation}
For the metric \Ref{firstorder.m} the last equation takes the
form:
\beq\label{HJ}
-e^{-2u}\left(\frac{\partial S}{\partial t}\right)^2+
e^{2u}\left(\frac{\partial S}{\partial r}\right)^2+
\frac{e^{2u}}{r^2+r_0^2}\left(\frac{\partial S}{\partial
\varphi}\right)^2- 2\omega e^{-2u}\frac{\partial S}{\partial
t}\frac{\partial S}{\partial \varphi}+\mu^2=0,
\eeq
where $\omega$ is given by Eq. \Ref{sol}. Since the metric does
not depend explicitly on time $t$ and the angle $\varphi$, we look
for a solution of the equation \Ref{HJ} in the following form
\begin{equation}\label{Action}
S=-Et+L\varphi+S_r(r),
\end{equation}
where $E$ is conserving energy, $E>\mu$, and $L$ is a projection
of angular momentum on $z$ axis. Substituting the expression
\Ref{Action} into \Ref{HJ} we find $S_r(r)$ and obtain the
solution
\beq\label{solS}
S=-Et+L\varphi+\int\left[e^{-4u}E^2-2\omega
e^{-4u}EL-\frac{L^2}{r^2+r_0^2}-e^{-2u}\mu^2\right]^{1/2}dr
\eeq
Using the equalities
$$
p^\alpha=\mu\frac{dx^\alpha}{ds}=g^{\alpha\beta}p_\beta=g^{\alpha\beta}\frac{\partial
S}{\partial x^{\beta}}
$$
we find the equations of motion explicitly:
\bea
\mu\frac{dt}{ds}&=&e^{-2u}E-\omega e^{-2u}L,\label{eqm1}\\
\mu\frac{dr}{ds}&=&\left[E^2-2\omega
EL-\frac{e^{4u}L^2}{r^2+r_{0}^2}-e^{2u}\mu^2\right]^{1/2},\label{eqm2}\\
\mu\frac{d\varphi}{ds}&=&\omega e^{-2u}E+
\frac{e^{2u}L}{r^2+r_{0}^2}.\label{eqm3}
\eea
Further for the sake of simplicity we restrict our analysis to the
case $L=0$, that corresponds to the radial motion of a test
particle at $r=+\infty$. In this case one finds from Eqs.
\Ref{eqm1} and \Ref{eqm3}
\beq
\frac{d\varphi}{dt}=\omega.
\eeq
This relation clearly demonstrates that $\omega$ is nothing but a
local angular velocity of a free falling test particle. The
particle's trajectory can be found from Eqs. \Ref{eqm2} and
\Ref{eqm3}:
\beq\label{tr}
\varphi-\varphi_0=\int\frac{\omega
e^{-2u}dr}{\left[1-e^{2u}(\mu/E)^2\right]^{1/2}},
\eeq
where the functions $u$ and $\omega$ are given by Eqs. \Ref{u} and
\Ref{sol}, respectively. Taking into account the asymptotical
properties of $u$ and $\omega$ we obtain from Eq. \Ref{tr} that
\beq\label{tr1}
\varphi-\varphi_0=0 \quad{\rm at}\quad r\to+\infty,
\eeq
and
\beq\label{tr2}
\varphi-\varphi_0=\frac{\omega_0e^{-2\pi m/r_0
}r}{{\left[1-e^{2\pi m/r_0}(\mu/E)^2\right]^{1/2}}} \quad{\rm
at}\quad r\to-\infty.
\eeq
The first relation \Ref{tr1} just illustrates that in the
asymptotical region $r=+\infty$ the particle's trajectory goes
along the radial direction with a constant $\varphi$. The second
relation \Ref{tr2} shows that after the particle has passed
through the rotating wormhole's throat and gone to the
asymptotical region $r=-\infty$ it receives a rotational momentum,
so that its trajectory represents a spiral with the constant step
$\Delta r$:
\beq
\Delta r=\frac{2\pi{\left[1-e^{2\pi
m/r_0}(\mu/E)^2\right]^{1/2}}}{\omega_0e^{-2\pi m/r_0 }}.
\eeq
Note that since $\omega_0\sim\alpha$ then $\Delta
r\sim\alpha^{-1}$, and so $\Delta r$ is large.

\begin{figure}[h]
\centerline{\includegraphics[width=6cm]{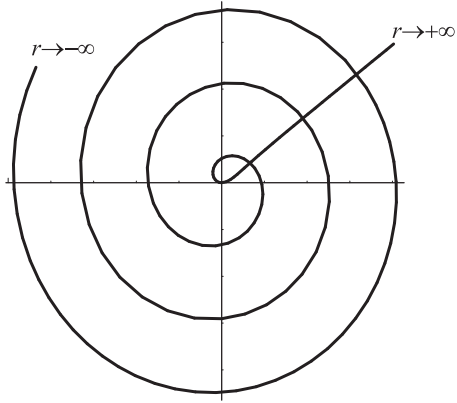}} \caption{The
trajectory of a test particle in the rotating wormhole spacetime.
\label{fig2}}
\end{figure}

The equation of trajectory becomes especially simple in the case
$m=0$ corresponding to the massless wormhole. In this case, using
Eq. \Ref{omegam0} and integrating, we find from \Ref{tr}
\beq\label{trm0}
\varphi-\varphi_0=\frac{\alpha r/r_0
}{\left[1-(\mu/E)^2\right]^{1/2}}
\left(1-\frac{2}{\pi}\arctan\frac{r}{r_0}\right).
\eeq
The trajectory \Ref{trm0} on the $(r,\phi)$-plane is shown in the
figure \ref{fig2}.

To study the propagation of light we also use the Hamilton-Jacobi
method \cite{Land}. For this aim we write the Hamilton-Jacobi
equation for the eikonal $\psi$ ($\mu=0$ for light):
\beq
g^{\alpha\beta}\frac{\partial\psi}{\partial
x^\alpha}\frac{\partial\psi}{\partial x^\beta}=0.
\eeq
As before we look for a solution in the form
\beq
\psi=-\vartheta t+L\varphi+\psi_r(r),
\eeq
where $\vartheta$ is the frequency of light observed at infinity.
Further calculations are similar to previous ones. In particular,
the trajectory of a light ray propagating in the region
$r=+\infty$ along the radial direction can be obtained directly
from Eq. \Ref{tr} merely setting $\mu=0$ and substituting
$\vartheta$ instead of $E$. As the result we find
\beq\label{trlight}
\varphi-\varphi_0=\int{\omega e^{-2u}dr}.
\eeq
The trajectory \Ref{trlight} is qualitatively similar to that
shown in Fig. \Ref{fig2}. Namely, the ray of light, propagating
initially along the radial direction, becomes ``twisted'' by the
wormhole rotation. This means that after the ray of light has
passed through the wormhole throat and gone to the region
$r=-\infty$, it becomes to be propagating along a spiral
trajectory with step $\Delta r=2\pi\omega_0^{-1}e^{2\pi m/r_0 }$.

\section{Summary and conclusions\label{conc}}   %
We have constructed and analyzed the solution describing the
rotating wormhole in the theory of gravity with the scalar field
with negative kinetic energy. The solution has been obtained under
the assumption about a slow rotation of the wormhole. In this case
there is the parameter of smallness which characterizes the
distinction of a rotating wormhole from a static one. Using this
parameter one can search for approximate solutions with respect to
the order of smallness. We have solved the problem in the
framework of the first order approximation. In this approach the
static spherically symmetric wormhole has been chosen as the zero
order solution. Solutions of next orders can be found as
respective perturbations of the zero order solution. Up to the
first order the only metric function $\omega$, i.e. the local
angular velocity of rotation, has obtained a small correction. The
other metric functions and the scalar field remain to be
unperturbed (see Eq. \Ref{firstorder.m}).

The analysis of motion of test particles in the rotating wormhole
spacetime reveals an interesting feature. The particle initially
propagating along the radial direction turns out to be involving
into the wormhole rotation so that after passing through the
throat of wormhole it continues its motion along a spiral
trajectory moving away from the throat. The similar behavior has
the propagation of light. The ray of light after passing through
the rotating wormhole throat is propagating along the spiral.

Note that a number of interesting and important problems remain
unsolved in the paper. Namely, how does the rotation of wormhole
tell on the value of violation of the null energy condition, i.e.
is it increasing or decreasing? Or, how does the rotation change
characteristics of the static wormhole such as the throat radius?
To answer these questions one should consider at least the second
order approximation. Such the consideration will be presented in
\cite{KasSus}.

\section*{Acknowledgments}
The work was supported by the Russian Foundation for Basic
Research grants No 05-02-17344, 05-02-39023.

\end{document}